\documentclass[12pt]{article}
\usepackage{epsfig}

\def\a{\alpha}

\def\b{\beta}

\def\th{\theta}
\def\Th{\Theta}

\def\s{\sigma}
\def\f{\phi}
\def\F{\Phi}

\def\e{\epsilon}
\def\r{\rho}
\def\ch{\chi}
\def\G{\Gamma}
\def\g{\gamma}
\def\o{\omega}

\def\d{\delta}
\def\m{\mu}
\def\n{\nu}
\def\l{\lambda}
\def\L{\Lambda}

\def\t{\tau}

\def\cl{{\cal L}}

\def\cv{{\cal V}}
\def\ca{{\cal A}}
\def\cb{{\cal B}}

\def\ch{{\cal H}}
\def\ha{\frac12}

\def\to{\rightarrow}

\newcommand{\be}{\begin{equation}}
\newcommand{\ee}{\end{equation}}
\newcommand{\bea}{\begin{eqnarray}}
\newcommand{\eea}{\end{eqnarray}}

\def\ket#1{\left|#1\right\rangle}

\begin{document}

\begin{titlepage}


\bigskip
\bigskip
\bigskip
\bigskip

\begin{center}

{\bf{\Large Spin Foam Models of Matter Coupled to Gravity }}

\end{center}
\bigskip
\begin{center}
 A. Mikovi\'c \footnote{E-mail address: amikovic@ulusofona.pt. An
 associate of Institute of Physics, P.O.Box 57, 11001 Belgrade, Yugoslavia}
\end{center}
\begin{center}
 Departamento de Matem\'atica e  Ci\^encias de
Computa\c{c}\~{a}o, Universidade Lusofona, Av. do Campo Grande
376, 1749-024 Lisboa, Portugal
\end{center}

\normalsize

\bigskip
\bigskip
\begin{center}
                        {\bf Abstract}
\end{center}
We construct a class of spin foam models describing matter coupled
to gravity, such that the gravitational sector is described by the
unitary irreducible representations of the appropriate symmetry
group, while the matter sector is described by the
finite-dimensional irreducible representations of that group. The
corresponding spin foam amplitudes in the four-dimensional gravity
case are expressed in terms of the spin network amplitudes for
pentagrams with additional external and internal matter edges. We
also give a quantum field theory formulation of the model, where
the matter degrees of freedom are described by spin network fields
carrying the indices from the appropriate group representation. In
the non-topological Lorentzian gravity case, we argue that the
matter representations should be appropriate $SO(3)$ or $SO(2)$
representations contained in a given Lorentz matter
representation, depending on whether one wants to describe a
massive or a massless matter field. The corresponding spin network
amplitudes are given as multiple integrals of propagators which
are matrix spherical functions.
\end{titlepage}
\newpage

\section{Introduction}

The Barrett-Crane spin foam state sum models of four-dimensional
quantum gravity \cite{bce,bcl,pre,prl}, represent a promising
approach for constructing a quantum theory of gravity. These are
non-topological state sum models, and they give discrete spacetime
three-geometry to three-geometry transition amplitudes \cite{mik}.
One can consider these models as examples of regularized path
integral approach to quantum gravity. The spacetime is described
by a simplical complex, and the state sum is over colored
triangulations. This approach is a category theory generalization
of the Regge calculus, where the basic category involved is the
category of representations of the appropriate symmetry group.
This group is taken to be $SO(d)$ in the $d$-dimensional Euclidian
gravity case, or $SO(d-1,1)$ in the Lorentzian case. Other groups,
like anti de-Sitter, can be considered as well \cite{mik}.

One can also think of the spin foam models as the
higher-dimensional generalizations of the string theory matrix
models, since the spin foam amplitudes are given by a field theory
Feynman diagrams which are dual to spacetime triangulations
\cite{boul,oo,dfkr,rr,mik}. In the $d=2$ case one recovers the
string theory matrix models \cite{mik,lpr}.

The standard way of making the spin foam partition function and
the amplitudes finite is by passing to the corresponding quantum
group category \cite{tv,cky,bcl}. However, it was discovered in
\cite{pre,prl} that even in the Lie group case, the BC model could
be made finite by changing the field theory propagator. It has
been proven that the BC Euclidian model is perturbatively finite
\cite{pez}, and recently this has been proven  for the BC
Lorentzian model \cite{cpr}, i.e. the amplitude for any
non-degenerate finite triangulation of the spacetime is finite.

In order to make a spin foam model a serious candidate for a
quantum theory of gravity, like string theory, one should be able
to incorporate matter. Mathematically this would mean enlarging
the category of representations, by adding a new set of
representations corresponding to the matter fields. The first
proposal of this kind was made by Crane \cite{cr}, who suggested a
tensor category of unitary representations for the quantum Lorentz
group. In this proposal a subset of the irreducible
representations (irreps) corresponds to the gravitational degrees
of freedom (DOF), while the rest corresponds to the matter DOF.
The corresponding state sum model is topological, which is
justified as an acceptable feature of the theory at high energies.
Then one conjectures that the local dynamics would appear at low
energies, by some sort of symmetry breaking mechanism. In a more
recent paper \cite{c2}, Crane has proposed to interpret the graphs
of the field theory formulation which do not correspond to
triangulations of manifolds, as triangulations of manifolds with
conical singularities, and then to interpret these singularities
as matter DOF. These are interesting and very geometrical
proposals; however, it is not clear how to  relate these matter
DOF to the particle field theory description in terms of fields
carrying pseudo-unitary finite-dimensional Lorentz irreps.

In this paper we give a formulation of a spin foam model with
matter inspired by the particle field theory formalism \cite{w}.
We consider a class of spin foam models describing matter coupled
to gravity, where the gravitational sector is described by the
unitary irreps of the appropriate symmetry group, while the matter
sector is described by the finite dimensional irreps of that
group. In the non-topological four-dimensional Lorentzian gravity
case, the gravitational sector will be described by the BC model,
while in the matter sector we will argue that one needs to take
appropriate $SO(3)$ or $SO(2)$ representations contained in a
given finite-dimensional Lorentz irrep. The construction we
present here naturally follows from the field theory formulation
of the spin foam amplitudes \cite{mik}. It was shown there that a
spin foam transition amplitude can be represented as an matrix
element of an evolution operator, where the initial and the final
states correspond to the three-dimensional spin networks (or spin
nets for short) induced by the boundary triangulations. One can
represent these spin net states as the spin net operators acting
on an appropriate Fock space vacuum. One can think of these spin
net operators as the second quantized spin net wave-functions from
the canonical loop gravity \cite{rs}. Since the spin net
wave-functions carrying fermionic matter have been introduced in
the context of canonical loop gravity \cite{mtr,bk}, it is not
difficult to generalize these results to the second quantization
formalism\footnote{The spin network quantum field theory is an
example of what is traditionally called a third quantization of
gravity, because it allows for more then one universe (a boundary
triangulation) in the formalism \cite{mik}. Since general
relativity is a classical field theory, then the expression
``second quantization" is a more appropriate terminology. However,
the exact relationship between the loop gravity formulation and
the spin-foam formulation is not yet fully understood.}, so that
the matter insertions at the spin net vertices become appropriate
creation/annihilation operators.

In section 2 we give basic definitions and formulas concerning the
spin net functions. In section 3 we give basics on spin foams, and
explain how the spin foam amplitudes are related to the spin net
amplitudes. In section 4 we construct the fermionic spin net
operators and states, and describe the general structure of the
interaction with the gravitational background. In section 5 we
discuss the fermions in the Euclidian gravity case. We show that
in the non-topological gravity case, the corresponding spin-net
amplitudes can be written as the multiple integrals of propagators
over the homogenous space $S^3 = SO(4)/SO(3)$, where the fermionic
propagator is given by a matrix zonal spherical function. These
results are straightforwardly generalized to the Lorentzian
gravity case in the next section. In section 7 we discuss the
general structure of the interaction terms for the case of matter
fields of arbitrary spins. In the non-topological gravity case,
the spin-net matter propagators are described by the corresponding
matrix spherical functions. We also argue that the spin-net
propagators for the analogues of the massless gauge bosons are the
matrix spherical functions restricted to an $SO(2)$ subgroup. In
section 8 we present our conclusions.

\section{Spin net functions and amplitudes}

We review the results on spin nets \cite{ba} from a point of view
which is most suitable for the type of generalizations we want to
make.

Consider square-integrable functions $\Psi$ on $G^E$, where $G$ is
a Lie group and $E$ is a natural number. A gauge invariant
function is determined by an oriented  graph $\G$ with $V$
vertices and $E$ edges, where a group element is associated to
each edge $e\in E$. If we label the vertices of $\G$ as $v_j$,
$1\le j \le V$, then the edges connecting the vertices $j$ and $k$
can be denoted as $e_{jk}$, and the corresponding group elements
as $g_{jk}$. Orientation is such that $g_{jk}=g^{-1}_{kj}$. A
function $\Psi_\G$ associated to the graph $\G$ is gauge invariant
if \be \Psi_\G(\,g_{jk}\,) = \Psi_\G(\,h_j \,g_{jk}\, h_k^{-1}\,)
\quad,\label{ginv}\ee for all $h_j , h_k \in G$.

By using the Peter-Weyl theorem, one can show that a gauge
invariant function $\Psi_\G$ can be expanded as \be \Psi_\G \,(g_1
,...,g_E ) = \sum_{\L,i}\, c_{\L,i}\, \F_{\G, \L , i}\,(g_1
,...,g_E)\ee where $\L = \{\L_1 , ..., \L_E \}$ denote the unitary
irreps of $G$ labeling the edges of $\G$, and $i =\{i_1 ,....,i_V
\}$ denote the intertwiners associated to the vertices of $\G$.
The basis functions $\F$ are called the spin net functions, and
they are given by  \be \prod_{e\in E}
{D^{(\L)}}_{\a_e}^{\b_e}(g_e)\prod_{v\in
V}C_{...}^{...\,(i_v)}\quad,\ee where $D^{(\L)}(g)$ is the
representation matrix of $g$, acting in the space $V(\L)$. All the
representation indices $\a_e$ and $\b_e$ are contracted by the
indices from the product of the intertwiner components.
Contraction is defined as $X^\a Y_\a = C^{\a\b}X_\a Y_\b$, where
$C^{\a\b}$ are the components of the $V(\L) \to V(\L^*)= V^* (\L)$
intertwiner, and $V^*$ is the dual vector space. Since any
intertwiner can be represented as a linear map
$V(\L_1)\otimes\cdots\otimes V(\L_n) \to \bf C$, we denote the
corresponding vector space as $Inv(\L_1 ,..., \L_n )$.

Given a spin net function $\F_{\G,\L,i}(g_e)$, one can obtain a
group invariant associated to the spin net \be A (\G,\L ,i)=
\F_{\G,\L,i}(g_1 = g_0 ,...,g_E =g_0 ) \quad,\ee where $g_0$ is
the identity element of $G$. We will call the number $A$ the
amplitude for the spin net $(\G,\L,i)$. These objects are the
basic building blocks of the state sum amplitudes. Since $D_\a^\b
(g_0) = \d_\a^\b$, the amplitude $A$ is simply the product of the
intertwiners $C^{...(i_v)}_{...}$ for all vertices $v\in V$, with
all the representation indices being contracted. The graphs $\G$
which are relevant for the state sum models are the one skeletons
of $d$-simplices, as well as the $\Theta_d$ graphs (two points
joined by $d$ edges).

In the case of open spin nets, the above results can be easily
generalized. The amplitude $A$ for an open spin net  will be a
group tensor given by the product of intertwiners for the
vertices, where all the representations indices are contracted,
except the indices corresponding to the irreps labelling the
external edges. This can be represented as \be
A_{\a_1...}^{\b_2...}\,(\G,\L,i) = C_{...}^{...\,(i_1)}\cdots
C_{...\a_1 ...}^{........\, (i_k)}\cdots C_{.........}^{...\b_2
...\,(i_l)}\cdots C_{...}^{...\,(i_V)}\quad.\ee

Hence the amplitude for an open spin net is not an invariant, but
it is a tensorial quantity. One can construct a group invariant by
contracting the external edge indices with the vectors and the
co-vectors from the external irrep vector spaces \be S
(\G,\L,i,v,u)= A_{\a_1...}^{\b_2 ...}\,(\G,\L,i) v_{\b_2} \cdots
u^{\a_1}\cdots \quad,\label{gta}\ee where $v_{\a}$ are the
components of a vector $v \in V(\L)$, while $u^{\a}$ are the
components of a covector $u \in V^*(\L)$. This formula will be
useful for constructing the invariant actions, since it is the
group theory description of the actions used in particle field
theories.

It is now also clear how to construct the open spin net functions
$\F_{\a_1 ...}^{\b_2 ...}\,(\,g_e \,)$. Note that the expansions
of the functions over $G^d$ used in the field theory
reformulations of the $d$-dimensional spin foam amplitudes as
Feynman graphs \cite{boul,oo,dfkr,rr,mik}, are simply expansions
in terms of open spin net functions associated to the $\times_d$
graphs ($d$ edges emanating from a single vertex).

In the case of the BC model \cite{bce,bcl}, the amplitude for the
$\times$ graph is not just the intertwiner $C^{(i)}_\times \in
Inv(\L_1,...,\L_4)$, but it is a linear combination of these \be
S_\times = \sum_{i} \a_i C^{(i)}_\times \quad.\ee This is the
consequence of the fact that the subcategory of irreps which are
used have invariant vectors under the $SO(3)$ subgroup.

\section{Spin foam amplitudes}

The state sum models which have been used for constructing
$d$-dimensional quantum gravity models, for a review see
\cite{ba,or}, are based on a two-complex $J=(V,E,F)$, consisting
of a finite number of vertices $v\in V$, edges $e \in E$ and faces
$f\in F$. The two complex $J$ is taken to be dual to a
triangulation of the spacetime manifold $M$. Since the faces of
$J$ are labeled with the irreps of $G$, one can regard this object
as a generalization of the spin net, where the one-complex $\G
=(V,E)$ is replaced by a set $(E,F)$ from the two complex $J$.
That is why $J$ is called a spin foam \cite{ba}. Hence these state
sum models give the amplitudes for the spin foams. An amplitude
for a spin foam can be written as \be A(J) = \sum_{\L} \prod_{f\in
F} A_{2} (\L_f) \prod_{e\in E} A_{1} (\L_{f_1 (e)}\cdots \L_{f_k
(e)})\prod_{v\in V}A_{0} (\L_{f_1 (v)}\cdots \L_{f_l (v)})
\quad,\label{sfa}\ee where $A_2$ is the amplitude for a face,
$A_1$ is the amplitude for an edge, and $A_0$ is the amplitude for
a vertex of the two complex $J$.

There are several ways to determine the amplitudes $A_n$, but what
is important for our purposes is the fact that these amplitudes
are proportional to the amplitudes for the closed spin nets
defined in the previous section. In the $d=4$ spacetime case,
$A_0$ is the amplitude for the 4-simplex graph (a pentagram)
\cite{oo,bce,bcl}. $A_1$ can be one, or it is the amplitude for
the $\Theta_4$ graph in the modified BC model \cite{pre,prl,cpr}.
$A_2$ is usually taken to be $dim\L$, which is a single loop spin
net amplitude. One can also take $A_2$ to be any function of $\L$,
such that a group Fourier transform of $A_2$ exists \cite{op,pf}.

Another aspect of the spin foam amplitude (\ref{sfa}) which will
be important is the fact that when $A_2 = dim\L$  then the
amplitude can be represented as a Feynman diagram of a field
theory over the group $G$ \cite{boul,oo,dfkr,rr,mik}. The action
for this field theory can be written in terms of the Fourier modes
$\f$, which transform like tensors from $Inv (\L_1,...,\L_d)$
spaces. In the $d=4$ case the action takes the form \bea S_4 &=&
\frac12 \,\sum_{\L,i} Q_{\vec\a_1
\vec\a_2}(\times_1, \times_2)\f^{\vec\a_1}_{\times_1} \f^{\vec\a_2}_{\times_2}\nonumber\\
&+& {1\over 5!}\sum_{\L ,i} \cv_{\vec\a_1 \cdots \vec\a_5}
(\times_1,\cdots ,\times_5)\,\f^{\vec\a_1}_{\times_1}\cdots
\f^{\vec\a_5}_{\times_5}\quad,\label{fta}\eea where
$\f^{\vec\a}_{\times} = \f^{\a_1 \cdots \a_4 }_{\L_1 \cdots \L_4
(i)}$. The vertex functions $Q$ and $\cv$ are given by \bea
Q^{-1}(\times_1, \times_2) &=&
\d^4\,A_1(\L_1,\cdots,\L_4 ;i_1 ,i_2)\\
\cv (\times_1,\cdots,\times_5) &=& \d^{10} A_0
(\L_1,\cdots,\L_{10};i_1 ,\,...\,,i_5)\quad,\eea where $\d$ are
the appropriate delta functions of the indices. $A_1$ can be the
amplitude for the $\Theta_4$ graph, or it could be one, when is
the amplitude for four parallel lines. $A_0$ is the amplitude for
the pentagram. In the case of BC models, $A_0$ and $A_1$ will not
depend on the intertwiner labels $i_k$.

The corresponding Feynman diagram is just the one complex $(V,E )$
obtained from the two complex $J=(V,E,F)$. This diagram is a
$(d+1)$-valent graph in the case of $d$-dimensional model. Its
value is given by (\ref{sfa}), which can be represented by a
collection of spin net diagrams, such that to each vertex we
associate a $d$-simplex graph, while to each edge we associate $d$
parallel lines, or a $\Theta_d$ graph in the modified version.

\section{Fermionic matter}

The Fourier modes $\f^{\vec\a}_{\times}$ can be promoted into
creation and annihilation operators \cite{mik}, so that the spin
foam amplitude can be represented as an matrix element of an
evolution operator, in analogy with the particle field theory
case.  As shown in \cite{mik}, one can construct the ``in'' and
the ``out'' states which describe the spatial spin nets
corresponding to the boundary triangulations of the spacetime
manifold. Such spin net states can be represented by \be
\ket{\g}=\prod_{v\in V(\g)} \f^\dag_{\times_v } \ket{0}\quad,
\label{sns}\ee where $\ket{0}$ is the vacuum of the Fock space
constructed from the creation and annihilation operators $\f$ and
$\f^\dag$, and $\g$ is a four-valent spin net graph dual to the
boundary triangulation. All the representation indices $\a_v$ in
(\ref{sns}) are appropriately contracted.

In the loop gravity formalism \cite{mtr,bk}, the fermionic matter
is introduced by replacing the spin net $\g$ with an open spin net
$\g_F$, which is obtained from $\g$ by putting an external edge
carrying a fermionic irrep $\L_s$ at each site of $\g$ where a
fermion is located. One can also put more than one fermion at each
lattice site, up to $dim \L_s$ (Pauli exclusion principle). In
that case, the external edge is labeled with an irrep from an
antisymmetrized tensor product of $k$ $\L_s$ irreps \cite{bk}.

In the spin foam formalism, the analogue of the loop gravity
construction would be to construct the $\ket{\g_F}$ state by
introducing the fermionic creation and annihilation operators
$\psi_\s (\times_v)$ and $\psi_\s^\dagger (\times_v)$, where $\s$
is the representation index of $\L_s$, and the label $\times_v$
denotes the four irreps of the spin net site where the fermion is
located. Hence the corresponding state would be given as \be
\ket{\g_F}=\prod_{v^\prime \in
V^\prime(\g)}\prod_{\s_{v^\prime}}\psi^\dag_{\s_{v^\prime}}(\times_{v^\prime})
\prod_{v\in V(\g)} \f^\dag_{\times_v}\ket{0} \,,\label{fsn}\ee
where $V^\prime$ is the set of vertices where the fermions are
located.

In order to calculate the transition amplitudes between two
fermionic spin net states, which would be a fermionic spin foam
amplitude, we need to introduce the corresponding terms in the
field theory action (\ref{fta}). Since classically the fermionic
propagation is described by a line in the spacetime, and since the
fermions interact with the gravitational field, the candidate
action will be \be S_F = \sum_{\L}
\psi^{\s}(\times)\psi^{\s^\prime}(\times^\prime ) \cv_{\s
\s^\prime} (\times ,\times^\prime , \times_1,\cdots, \times_n )
\f_{\times_1} \cdots \f_{\times_n} + (h.c.)\quad,\ee where
$(h.c.)$ stands for the hermitian conjugate term.

The form of the vertex $\cv$ can be restricted by two
requirements. First, $S_F$ has to be a group invariant, i.e. a
scalar. Second, we require that the fermions propagate only on
complexes which are dual to the Feynman graphs of the
gravitational model (\ref{fta}). The idea behind this is that the
purely gravitational model builds up the space-time on which the
fermions propagate. Hence $n=5$, and \be \cv_{\s \s^\prime}
(\times ,\times^\prime , \times_1,\cdots, \times_5 ) = \d^{10}
\d_{\times,\times_1} \d_{\times^\prime,\times_k} A_{\s
\s^\prime}(\L_1,...,\L_{10})\,,\ee where $A_{\s \s^\prime}$ is a
spin net amplitude for the pentagram with two external edges
attached at site $1$ and site $k$, where $1\le k \le 5$.

Note that one can take a more complicated spin net graph for the
amplitude $A_{\s\s^\prime}$, by adding internal edges labelled by
the matter irrep, see sect. 5. One would then need to study the
semiclassical limit in order to determine which spin net would
give a correct physical theory. Since that is beyond the scope of
this paper, and for the sake of simplicity, we will analyze only
the simplest possible spin nets.

We will also add to $S_F$ a purely quadratic fermion term \be
\sum_\L \psi^\s (\times)Q_{\s \s^\prime}(\times ,\times^\prime )
\psi^{\s^\prime} (\times^\prime) + (h.c.)\quad,\ee in order to
have a well-defined perturbative expansion. The propagator
$Q^{-1}$ will be determined by the amplitude for the $\Th_4$ graph
with two external edges labelled with $\L_s$, or it could be a
trivial amplitude for five parallel lines. We denote the
propagator amplitude as $G_{\s \s^\prime}$.

Note that $(\psi_\s)^* = \psi^\s =
C^{\s\s^\prime}\psi_{\s^\prime}$, and the vector space duality $*$
is in general different from the complex conjugation (reality)
properties. The reality properties determine relation between the
creation and annihilation operators $\psi$ and $\psi^\dagger$
\cite{mik}.

Therefore the maximal closed spin net subgraph for
$A_{\s\s^\prime}$ has to be at least the pentagram graph. Hence a
relevant fermionic Feynman diagram $\G_F$ which is generated by
the $S_4 + S_F$ action will be determined by a spacetime skeleton
diagram $\G$ (a 5-valent graph dual to a spacetime triangulation)
plus a line connecting the vertices of $\G$. Therefore $\G_F$ will
be a graph consisting of 5-valent and 7-valent vertices, see Fig.
1. Out of these diagrams we will consider only those for which the
fermion path is a line which connects the centers of a string of
adjacent 4-simplices, extending from the initial to the final
boundary.

Hence the fermionic spin foam $J_F$ will be given by the usual
spin foam $J$ and a line $L$ of edges starting from an ``in" edge
and terminating with an ``out" edge. The corresponding amplitude
will be given by \be A_{\s_i}^{\s_f}(J_F) =  \sum_{\L}\prod_{f\in
F} A_f (\L) A_{\s_i}^{\s_f}(L)\prod_{e \in E^\prime}A_e(\L)
 \prod_{v \in V^\prime} A_v(\L) \,,\label{gsfa}\ee
where \be  A_{\s_i}^{\s_f}(L)= A_{\s_i}^{\s_1}(v_1)\,
G_{\s_1}^{\s_2}(e_{12})\,A_{\s_2}^{\s_3}(v_2)\, \cdots
\,G_{\s_{k-1}}^{\s_k}(e_{k-1,k})\,A_{\s_k}^{\s_f}(v_k)\quad.
\label{fba}\ee

The number of the ``in" and the ``out" fermions has to be the same
for a free theory, and the above formula can be then easily
generalized to the case when there are several lines $L$.

The amplitude $A$  for a fermionic vertex of $J_F$ will be given
by a spin net amplitude for a pentagram with two external
fermionic edges, Fig. 2 \be A_{\s^\prime}^\s (\L_1,\,
...\,,\L_{10})= \sum_\a C_{\s^\prime \times}C^\s_{\times}
C_{\times}C_{\times}C_{\times}\quad, \label{mina}\ee where
$C_{\s^\prime \times},C^\s_{\times}\in Inv(\L_s , \L_1,...,\L_4)$
and $C_{\times}\in Inv(\L_1,...,\L_4)$. The amplitude for an
fermionic edge of $J_F$ will be given by the propagator amplitude
$G_\s^{\s^\prime}$, which in the non-trivial case is given by the
amplitude for the $\Theta_4$ graph with two external edges, Fig.
3\be G_{\s^\prime}^\s (\L_1,\, ...\,,\L_{4})= \sum_\a C_{\s^\prime
\times}C^\s_{\times} \quad. \label{minp}\ee

In some of the cases the minimal amplitudes (\ref{mina}) and
(\ref{minp}) are zero, because $C_{\s\times} =0$. Then one can
replace it with $C_{\s\s^\prime \times}$ from $Inv(\L_s,\L_s ,
\L_1,\,...\,,\L_4)$, so that \be A_{\s^\prime}^\s (\L_1,\,
...\,,\L_{10})= \sum_\a C_{\s^\prime \times}^\t C_{\t \times}^\s
C_{\times}C_{\times}C_{\times}\quad.\ee The corresponding spin net
graph is a pentagram with two external and one internal fermion
edge, joined in a line, Fig. 4. Similarly, for the edge amplitude
$G_\s^{\s^\prime}$, we will have in the non-trivial case the
$\Th_5$ graph with two external edges, Fig. 5, so that \be
G_{\s^\prime}^\s (\L_1,\, ...\,,\L_{4})= \sum_\a C_{\s^\prime
\t\times}C^{\t\s}_{\times} \quad. \ee

\section{Euclidian case}

We now study the four-dimensional Euclidian gravity spin foam
model \cite{bce,pre}, so that $G=SO(4)$. The irreps of $SO(4)$ can
be labeled as $\L = (j,k)$ where $j$ and $k$ are the half-integers
labelling the irreps of $SU(2)$. In the topological gravity case,
one uses all $SO(4)$ irreps for labelling the triangles of a
triangulation of $M$. The fermions will be represented by $\L_s^+
= (\ha , 0) $ and $\L_s^- =(0,\ha)$ irreps. These irreps
correspond to chiral, or Weyl fermions, while a Dirac fermion
would be a reducible representation $(\ha,0) \oplus (0,\ha)$.

The relevant intertwiners  for constructing the fermionic spin
foam amplitude are from the space $Inv(\L_s,\L_1 ,...,\L_4)$.  A
basis $\ket{i}$ is determined by \be C^{\L_s \L_1  ... \L_4
\,(i)}_{\s \a_1 ...\a_4} = \sum_{\a,\a^\prime} C^{\L_s \L_1
\L}_{\s \a_1 \a} C^{\L \L_2\L^{\prime}}_{\a \a_2
\a^\prime}C^{\L^{\prime} \L_3 \L_4}_{\a^\prime \a_3
\a_4}\quad.\label{ifive} \ee In order to simplify the notation, we
take $X_\a Y_\a = X^\a Y_\a = C^{\a\b}X_\a Y_\b$. Note that the
formula (\ref{ifive}) can be represented graphically by Fig. 6.
Hence the intertwiner label is given by $i=(\L,\L^\prime)$, and we
can now construct the fermionic amplitudes for topological gravity
spin foam model by using the procedure described in sect. 4.
However, we will not discuss further the topological gravity case,
and we will concentrate on the non-topological model, because of
its importance for physics.

The set of relevant irreps in the non-topological gravity case is
given by the simple irreps $N=(j,j)$. In that case $C^{\L_s N_1
... N_4 }_{\,\,......}=0$, which can be proven by using a $Z_2$
grading of the $SO(4)$ irreps. This grading is introduced by
splitting the irreps into even and odd, or bosonic and fermionic
irreps via the parity function $p\,(n)=\{-1,1\}$ if $n$ is odd,
even, respectively, as
 \be
p\,(j,k) = p\,(2j) p\,(2k)\quad. \ee Hence the simple irreps are
even, i.e. bosonic, while $\L_s^\pm$ are odd, i.e. fermionic.
Since $p\,(\L_1 \otimes \L_2) = p\,(\L_1) p\,(\L_2)= p(\L)$ where
$\L$ is an irrep from the tensor product, then a product of
bosonic irreps can never yield a fermionic irrep, and hence
$C^{\L_s  N_1 ... N_4 }=0$. Therefore the minimal option for the
intertwiner space is $Inv(\L_s ,\L_s , N_1 ,... ,N_4 )$, and a
basis $\ket{i}$ can be constructed from \be C^{\L_s \L_s N_1 ...
N_4 \,(i)}_{\s \s^\prime \a_1 ... \a_4} = \sum_{\b_1,\b_2,\b_3}
C^{\L_s \L_s B_1}_{\s\s^\prime \b_1}C^{N_1 N_2 B_2}_{\a_1
\a_2\b_2}C^{B_1 B_2 B_3}_{\b_1\b_2\b_3} C^{B_3 N_3
N_4}_{\b_3\a_3\a_4} \label{tfi}\ee where $B$ are the bosonic
irreps. The equation (\ref{tfi}) can be graphically represented by
Fig. 7.

Now we have to impose the constraints coming from the fact that
the simple irrep vector spaces have invariant vectors under the
$SO(3) \cong SU(2)$ subgroup. This amounts to constructing an
invariant vector in the corresponding $Inv(n)$ space. When there
are no fermions, then $n=4$, and the invariant vector, or the BC
vertex, is constructed as \cite{bce} \be S^{N_1 ... N_4 }_{\a_1
...\a_4} = \sum_{M} C^{N_1... N_4\,(M)}_{\a_1 ... \a_4} = \sum_M
\sum_\a C^{N_1 N_2 M}_{\a_1 \a_2 \a}C^{M N_3 N_4}_{\a \a_3 \a_4}
\,,\label{bcv}\ee where $M$ are the simple irreps. The $3j$
symbols for $SO(4)$ can be expressed as the product of $3j$
symbols for $SU(2)$ as \be C^{N_1 N_2 N_3}_{\a_1 \a_2 \a_3} =
C^{j_1 j_2 j_3}_{m_1 m_2 m_3}C^{j_1 j_2 j_3}_{n_1 n_2
n_3}\quad.\ee

The consequence of using the $BC$ vertex for constructing the spin
net amplitudes is that these amplitudes can be represented as
multiple integrals over the coset space $X=G/H$, which is a
three-sphere $S^3=SO(4)/SO(3)$ \cite{bar,fkb}. For example, the
amplitude for the $\Theta_4$ graph will be given by \be \Theta
(N_1 ,...,N_4) = \sum_\a S^{N_1 ... N_4 }_{\a_1 ...\a_4}{ S}^{N_1
... N_4 }_{\a_1 ...\a_4} \quad.\ee One can show that \be \Theta
(N_1 ,...,N_4) = \int_{X^2} dx dy \prod_{i=1}^4 K_{N_i} (x,y)
\quad,\ee where \be K_N (x, y) = Tr\, D^{(j)}(x^{-1}\cdot y)=
{\sin((2j+1)\th)\over \sin\th}\quad,\label{escp}\ee and $\th$ is
the geodesic distance between the coset points. This formula works
because the coset points $x$ and $y$ can be considered as $SU(2)$
group elements due to equivalence of $S^3$ and $SU(2)$ spaces.

We will now use this property to construct the fermionic BC
vertex. Consider the following integral associated to the
$\Theta_4$ graph \be \Theta_{ss^\prime} (N_1 ,...,N_4) =
\int_{X^2} dx dy K_{ss^\prime}(x, y)\prod_{i=1}^4 K_{N_i} (x,y)
\quad,\label{fti}\ee where \be K_{ss^\prime} (x, y) =
D^{(1/2)}_{ss^\prime}(x^{-1}\cdot y)\quad.\label{sp}\ee

One can now show that \be \Theta_{ss^\prime} (N_1 ,...,N_4) =
\sum_{\a,\tilde s} S^{N_1 ... N_4 N_{1/2}}_{\a_1 ...\a_4 \tilde s
s}{ S}^{N_1  ... N_4 N_{1/2}}_{\a_1 ...\a_4 \tilde s s^\prime }
\quad,\label{ftv}\ee where $S$ is an element of $Inv ( N_1
,...,N_4 , N_{1/2} )$ given by \be S^{N_1  ... N_4 N_{1/2}}_{\a_1
...\a_4 \s} = \sum_{M,N}
 \sum_{\a,\b} C^{N_1
N_2 M}_{\a_1 \a_2 \a}C^{M N_{1/2} N}_{\a \s \b} C^{N N_3 N_4}_{\b
\a_3 \a_4} \quad,\label{sv}\ee and $N_{1/2}=(\ha,\ha)$.

When (\ref{sv}) is compared to (\ref{tfi}), then it is easy to see
that the expression (\ref{sv}) can be also taken as an intertwiner
from the space $Inv(\L_s^+ , \L_s^- , N_1 ,...,N_4)$, due to
$N_{1/2} = \L_s^+ \otimes \L_s^-$. Therefore the expression
(\ref{fti}) can be interpreted as the amplitude for the open spin
net consisting of the $\Th_4$ graph plus two external edges
labelled with $\L_s^+$ and one internal edge labeled with
$\L_s^-$, Fig. 5.

Hence the action $S_F$ for the $\L_s$ fermions will be given by
\bea S_F &=& \sum_N \psi^\s_\pm (\times) Q_{\s\n}
(\times,\times^\prime)\psi_\pm^\n (\times^\prime)\nonumber\\ &+&
\sum_N \psi^\s_\pm (\times) \psi^\n_\pm (\times^\prime)\cv_{\s\n}
(\times,\times^\prime;\times_1 \cdots \times_5)
\f_{\times_1}\cdots \f_{\times_5} \nonumber\\ &+&
(h.c.)\quad,\label{fqa}\eea where $Q^{-1}$ and $\cv$ are
determined by the \be G_{\s\n}(N_1 ... N_4) = \int_{X^2}d^2 x
\,K_{\s\n} (x_1 , x_2) \prod_{i} K_{N_i} (x_1,x_2 )\label{edga}\ee
and \be A_{\s\n}(N_1 ... N_{10}) = \int_{X^5}d^5 x \, K_{\s\n}
(x_1 , x_k) \prod_{i<j} K_{N(ij)} (x_i ,x_j )\label{verta}\ee spin
net amplitudes, respectively. The edge amplitude (\ref{edga}) is
determined by the graph from Fig. 5, and the vertex amplitude
(\ref{verta}) is determined by the graph from Fig. 4.

The reason why (\ref{fti}) and (\ref{ftv}) are the same is that
the spinor propagator (\ref{sp}) is an example of a matrix
spherical function \cite{ch}, which is a generalization of the
scalar propagator $K_N$ (\ref{escp}). In general case \cite{ch},
if $H$ is a subgroup of $G$ and if $\t$ is an irrep of $H$,
consider irreps $\L$ of $G$ which contain $\t$ when decomposed
with respect to the subgroup $H$, i.e. \be V_\L = \bigoplus_{1\le
k\le n} V^k_\t \oplus ...= H_\t \oplus ...\quad.\ee Let $\F (g) =
P_\t D^{(\L)}(g)P_\t$, where $P_\t$ is the projector from $V_\L$
to $H_\t$. $\F (g)$ are called $\t$-spherical functions, and
$P_\t^k D^{(\L)}(g)$ are eigenfunctions of the Laplacian on the
homogenous vector bundle $E_\t =(G/H,V_\t )$. Since $\F (g) \in
Hom (H_\t , H_\t)$, we can construct $\f(g) = tr\,\F (g)\in
Hom(V_\t ,V_\t)$, where $tr$ is a partial trace over the
degeneracy label $k$. Then consider the object \be K_{\L,\t}(x,y)
= U(x,x_0) \f (g_x^{-1} g_y ) U (x_0 , y) \quad,\label{mp}\ee
where $U$'s are vector bundle parallel transport operators along
the geodesics $(x_0 , x)$ and $(x_0 , y)$, and $g_x = \s (x)$ is a
local section of the principal bundle $(G/H,G)$. By using $g_x =
xh$, $g_y = y \tilde h$, $U (x,x_0) = D^{(\t)} (h)$ and $U (x_0
,y) = D^{(\t)} (\tilde h^{-1})$, one can show that $K$ is a
function of only the coset space points. Note that when $\t$ is
the identity irrep, the expression (\ref{mp}) reduces to the
scalar propagator $K_N$, where $N$ is the class one irrep. Hence
(\ref{mp}) is the matrix generalization of $K_N$, and we can use
it to construct the spin net amplitudes for graphs with external
edges.

In the $SO(4)$ case, let $\vec\ca$ and $\vec\cb$ be the generators
of the two commuting $SU(2)$. Then the $SO(4)$ rotations
generators $J_{\m\n}$ can be expressed as\be \vec J =  \vec\ca +
\vec\cb
 \quad,\quad \vec K = \vec\ca - \vec\cb
\quad,\label{jk}\ee where $J_i = {\e\sb{i}}\sp{jk}J_{jk}$ and $K_i
= J_{i0}$. The $SO(4)$ irreps $\L$ are then determined by the
pairs of half-integers $(A,B)$, where $A$ and $B$ are the angular
momenta numbers of the two $SU(2)$. We then have \bea
D^{(A,B)}(\vec\th , \vec v) &=& \exp ( i \vec\th \vec J_\L + i
\vec v \vec K_\L )\nonumber\\&=& \exp (i \vec\th_+ \vec\ca_A + i
\vec\th_- \vec\cb_B )\nonumber\\ &=& D^{(A)} (\vec\th_+ ) \otimes
D^{(B)} (\vec\th_-) \,,\label{df}\eea where \be \vec\th_\pm = \vec
\th \pm \vec v \quad. \ee

The diagonal $SU(2)$ subgroup vectors are given by \be \ket{j, m}
= C^{jAB}_{mab} \ket{a}\otimes \ket{b} \quad. \ee These vectors
form an irrep space of the spatial rotation generators $\vec J$.
In the case of $\L_s$, we take $\t=j=1/2$, so that $P_\t = 1$.
From (\ref{df}) we get \be K^E_{1/2} (x,y) = D^{(1/2)} (x^{-1}
\cdot y)= D^{(1/2)}(-\vec v)D^{(1/2)}(\vec u )
\quad,\label{esp}\ee where \be D^{(1/2)}(\vec r) = \cos (r/2) I_2
+ i\,\vec\s \vec n \, \sin (r/2) \quad,\quad \vec n = {\vec r
\over \sqrt{(\vec r)^2}}= {\vec r \over r } \quad,\ee and $\vec\s$
are the Pauli matrices.

\section{Lorentzian case}

In the Lorentzian case $G=SO(3,1)\cong SL(2,C)$, and the
finite-dimensional irreps $(j,k)$ become pseudo unitary, because
the relation (\ref{jk}) becomes \be \vec J =  \vec\ca + \vec\cb
\quad,\quad \vec K = (- i) (\vec\ca - \vec\cb )
\quad,\label{ljk}\ee so that the boost generator $\vec K$ becomes
an anti-hermitian operator. This is the reason why there are no
finite-dimensional unitary irreps for the Lorentz group.

Hence the unitary Lorentz irreps are infinite-dimensional, and can
be labelled as $(m,\r)$, where $m$ is a half-intiger, while $\r$
is a non-negative real number \cite{gn}. The gravitational DOF are
again described by the simple irreps, or  the class one irreps
with respect to the spatial $SU(2)$ subgroup. These are given by
the $(m,0)$ and $(0,\r)$ irreps \cite{bcl}. However, only the
$(0,\r)$ irreps are used \cite{bcl}. We could then use the pseudo
unitary irreps $(j,k)$ to describe the matter fields, but we have
to check first the consistency of the state-sum model with unitary
and pseudo unitary irreps of the Lorentz group.

We now have $N=(0,\r)$ in the gravitational sector. The
corresponding propagator is given by \cite{bcl}\be K_\r (x,y) =
{\sin (\r d(x,y))\over \r \sinh d(x,y)} \label{lscp} \quad,\ee
where $d(x,y)$ is the geodesic distance between the points $x$ and
$y$ of the coset space $X$, which is now a hyperboloid
$SO(1,3)/SO(3)$. The fermions will carry the irreps $\L_s^\pm$.
The tensor product decomposition rule of a unitary and a
pseudo-unitary irrep can be inferred from the decomposition of
these irreps under the $SU(2)$ subgroup. For the $(m,\r)$ irrep we
have \cite{ruhl} \be \ch_{(m,\r)} = \bigoplus_{j=|m|}^\infty V_j
\quad.\ee Since $V(\L_s^\pm) = V_\ha$, we obtain  \be
\ch_{(m,\r)}\otimes V (\L_s^\pm) =
\ch_{(m-\ha,\r)}\oplus\ch_{(m+\ha,\r)}\quad,\ee which agrees with
\cite{cr}. This implies \be \ch_{(m,\r)}\otimes V (\L_v) =
\ch_{(m-1,\r)} \oplus\ch_{(m,\r)}\oplus\ch_{(m,\r)}
\oplus\ch_{(m+1,\r)}\quad,\ee where $\L_v = (\ha,\ha)$.

Hence $C^{\r_1 \r_2 \L_v}_{\,........} \ne 0$, and we can now
construct the Lorentzian analog of the BC spinor vertex
(\ref{sv}). For this we need a Lorentzian spinor propagator, which
is determined by the formula (\ref{mp}). It is given by an
analytic continuation of the Euclidian formula (\ref{esp}), where
$\vec v \to -i \vec v$, so that \be K_{1/2}^L (x,y) = D^{(1/2)}
(x^{-1} \cdot y)= D^{(1/2)}(i\vec v)D^{(1/2)}(-i\vec u ) \quad,\ee
where now \be D^{(1/2)}(i\vec r) = \cosh (r/2) I_2 + \vec\s \vec n
\, \sinh (r/2) \quad.\ee

\section{Interactions and gauge fields}

Note that in the discussion in section 4 one could have taken
$\L_s$ to be an arbitrary irrep of $G$, and moreover, one can take
a set of different irreps $\L_{s_1}=S_1$,...,$ \L_{s_k}=S_k$ to
represent fields of different spins. In the topological gravity
case, one can write the action as \bea S_I &=& \sum_{S} \sum_\L
\psi_S(\times)\psi_{S}(\times^\prime )Q_{SS}
(\times,\times^\prime)\nonumber\\ &+& \sum_S \sum_\L \psi_S
(\times) \psi_S (\times^\prime) \cv_{SS} (\times ,\times^\prime ,
\times_1,\cdots,
\times_5 ) \f_{\times_1} \cdots \f_{\times_5}\nonumber\\
&+&\sum_S \sum_\L \psi_{S_1}(\times) \cdots
\psi_{S_k}(\times^{(k)}) \cv_{S_1 ... S_k}
(\times,...,\times^{(k)}; \times_1 ,...,\times_5) \f_{\times_1}
\cdots \f_{\times_5} \nonumber\\ &+& (h.c.)\quad,\label{tia}\eea
where the vertex $Q$ will be determined by the spin net amplitudes
for the open graphs based on the five parallel lines and the
$\Th_4$, while the vertex $\cv$ will be determined by the spin net
amplitudes for open graphs based on the pentagram. A new feature
is that the set of possible graphs for the vertex $\cv_{S_1 ...
S_k}$ will contain graphs with more than five vertices, Fig. 8.

In the non-topological gravity case, we need to implement the $H$
invariance. This requires a specification of the irreps $s$ of $H$
which are contained in the irreps $S$ of $G$. In that case the
matter fields will have the indices from the $H$ irreps, and the
action can be written as \bea S_I &=& \sum_{S} \sum_N
\psi_s(\times)\psi_{s}(\times^\prime )Q^S_{ss}
(\times,\times^\prime)\nonumber\\ &+& \sum_S \sum_N \psi_s
(\times)\psi_s (\times^\prime )\cv^S_{ss} (\times ,\times^\prime ,
\times_1,\cdots,
\times_5 ) \f_{\times_1} \cdots \f_{\times_5}\nonumber\\
&+&\sum_S \sum_N \psi_{s_1}(\times) \cdots
\psi_{s_k}(\times^{(k)}) \cv^{S_1 ... S_k}_{s_1 ... s_k}
(\times,...,\times^{(k)}; \times_1
,...,\times_5) \f_{\times_1} \cdots \f_{\times_5} \nonumber\\
&+& (h.c)\quad,\label{ntia}\eea where the vertices $Q$ and $\cv$
will be determined by the spin net amplitudes for open graphs
based on five parallel lines, the $\Th_4$ and the pentagram
graphs, as in the topological case. However, these amplitudes will
be now given as multiple integrals of the propagators $K_N (x,y)$
and $K_{S,s} (x,y)$. Also note that the fields $\psi_s$ will now
carry the indices from the irreps of the subgroup $H$.

In the Euclidian/Lorentzian gravity case $H=SO(3)$, and it will be
instructive to study the case of spin $j=1$ matter fields, because
in the nature the fermions interact through the spin $j=1$ gauge
bosons. If we take a vector irrep $S_1 = (\ha,\ha)$, it will
contain a $s=j=1$ irrep of $SO(3)$. We denote the corresponding
spin net field as $A_i (\times)$, while the fermions we denote as
$\psi_{\s}^\pm (\times)=\psi_{s}^\pm (\times)$, where the
subscripts $\pm$ denote the corresponding $SO(4)$ irreps. The spin
foam analog of the particle field theory interaction term $\int
d^4 x \bar\psi \g^\m \psi A_\m$ can be written as \be \sum_N
\psi^+_s (\times)\psi^-_{s^\prime} (\times^\prime)
A_i(\times^{\prime\prime}) \cv_{+-}^{s \,s^\prime\,i}
(\times,\times^\prime,\times^{\prime\prime}; \times_1
,...,\times_5) \f_{\times_1} \cdots \f_{\times_5} + (h.c.)
\,,\label{uone}\ee where the vertex $\cv_{+-}$ is determined by
the spin net amplitude for the open graph given by the pentagram
plus a three-valent matter vertex, Fig. 9. This amplitude is given
by \be A_{+-}^{s \,s^\prime\,i} = \int_{X^6} d^5 x dy
K_\ha^{st}(x_1 ,y) K_{(\ha,\ha),1}^{ij}(x_4 ,y)K_\ha^{s^\prime
t^\prime} (x_2 ,y) C_{\,t \, t^\prime \, j }^{\ha\, \ha\,
1}\prod_{k<l} K_{N_{kl}}(x_k ,x_l)\,.\label{usna}\ee

Since (\ref{uone}) is a spin foam generalization of the particle
gauge field theory $U(1)$ interaction, it is natural to ask what
is the analog of the $U(1)$ gauge symmetry. One way to deal with
the issue of gauge symmetry is to assume that the components of
the fields appearing in the action (\ref{ntia}) are the physical,
or the independent DOF. Note that one can adopt this type of
approach in the context of flat space particle field theory
\cite{w}. One simply takes the creation and annihilation operators
labelled by the irreducible Poincare representation indices and
then forms polynomial interaction terms via the formula
(\ref{gta}). The relevant  Poincare irreps are the massive and the
massless ones, so that the index labels are the three-momentum,
the $SO(3)$ irrep indices for the massive case, or the
two-dimensional $Spin(2)\cong SO(2)$ representation labels in the
massless case.

Since one wants the particle interactions to be local in
spacetime, one then constructs spacetime fields carrying the
appropriate Lorentz irreps from the Poncare creation and
annihilation (c/a) operators, and writes the interactions as
spacetime integrals over the polynomials of these fields \cite{w}.
The problem with this approach is that the massless fields cannot
be always expressed in terms of the helicity c/a operators.  This
happens for the vector field $A_\m (x)$, which when expressed in
terms of the helicity c/a operators transforms non-homogenously
under the Lorentz transformations \cite{w}. However, the field
$f_{\m\n}=
\partial_{[\m} A_{\n]}$ transforms like a Lorentz tensor, so that
the cubic interaction term would be given by $\int d^4 x\bar\psi
\g^{\m\n}\psi f_{\m\n}$. But this is not the type of interaction
realized in nature, which is $\int d^4 x\bar\psi \g_\m \psi A^\m$.
This paradox is resolved by introducing the $U(1)$ gauge
transformations \cite{w}.

In the case of our spin foam model, this type of problem will not
appear immediately, because one is dealing only with the c/a
operators for the Lorentz group and its subgroups. These operators
carry the labels $\times = (N_1 ... N_4)$ and $SU(2)\cong SO(3)$
representation indices. There are no Poincare irreps labels, and
the reason for that is that the particle momentum is not conserved
in curved spacetimes. Hence it is not clear what would be a
physical interpretation of a spin-foam model based on the Poincare
group\footnote{This can be seen on the example of a Poincare spin
foam model based on the simple unitary irreps of the Lorentz
group. At first sight this appears a natural thing to do, because
the unitary irreps of the Lorentz group are also unitary irreps of
the Poincare group, since the Lorentz subgroup is the little group
for the zero momentum. Hence the gravitational sector would be
described by the zero-momentum unitary irreps, while the matter
sector would be described by the usual particle unitary irreps,
having a non-zero momentum. However, because the tensor product
conserves the momentum, the corresponding spin-foam amplitude
would conserve the momentum of a particle propagating in a curved
background.}

Our assumption that the components of the spin net fields are the
physical components can be further justified by the fact that
these fields live on the spin nets defined by the triangulations
of three-space boundaries, see equation (\ref{fsn}). We will then
consider these fields as the discretized analogs of the reduced
phase space canonical formalism fields. The canonical analysis of
the continuous gravity and matter actions shows that the effect of
the local Lorentz and the spacetime diffeomorphism constraints is
to reduce the number of components of the gravitational field,
while the matter field components are affected only by the gauge
symmetry constraints. The final result is that the non-gauge
fields have independent components which transform as the
appropriate representations of $SO(3)$, while the gauge fields
independent components transform as the appropriate
representations of $SO(2)$. In the flat space limit, these
independent component fields become the Fourier transforms of the
Poincare irreps, so that the non-gauge fields correspond to the
massive irreps, while the gauge fields correspond to the massless
ones\footnote{Except in the case of massless spin-half fermions,
or if there is a Higgs scalar field.}.

Therefore within this approach the $A_i (\times)$ operator cannot
be considered as a discrete massless gauge field analog, since it
carries a $j=1$ irrep of $SO(3)$, which corresponds to a massive
Poincare irrep. In order to introduce a massless-like vector gauge
boson in the model, we need a spin net field which transforms as a
two-dimensional representation of $SO(2)$. Therefore we need to
construct invariant propagators under an $SO(2)$ subgroup of the
$SO(3)$ subgroup.

Hence consider a subgroup $H^\prime$ of $H$, and let $\t^\prime$
be a representation (could be reducible) of $H^\prime$ contained
in the irrep $\t$ of $H$. Note that one cannot use the
$K_{\L,\t^\prime}$ propagators for this purpose, where $\t^\prime$
is contained in the irrep $\L$. The reason is that
$K_{\L,\t^\prime}$ will be a function on the coset space $X^\prime
= G/H^\prime$, which is different from the  space $X=G/H$, where
the gravitational propagators $K_N$ are defined. Therefore these
propagators can not be multiplied and integrated to give the spin
net amplitudes.

Note that the representation $\t^\prime$ will induce a vector
sub-bundle $(G/H ,V_{\t^\prime})$ of the vector bundle $(G/H
,V_{\t})$. We can then define the propagator \bea \tilde
K_{\L,\t^\prime}(x,y) &=& P_{\t^\prime}U(x,x_0) \f (g_x^{-1} g_y )
U (x_0 , y)P_{\t^\prime}\\&=&P_{\t^\prime}K_{\L,\t}(x,y)
P_{\t^\prime} \quad, \label{mpm}\eea where $P_{\t^\prime}$ is a
projector from $V_\t$ to $V_{\t^\prime}$. It is clear that $\tilde
K \in Hom(V^\prime,V^\prime)$ and it transforms correctly under $x
\to x h^\prime$ and $y \to y {\tilde h}^\prime$, where $h^\prime ,
\tilde h^\prime \in H^\prime$.

In our case we take the index on the massless-like spin net field
of spin $j$ to be from a reducible $Spin(2)$ helicity
representation \be V^\prime_j = \cl \{\ket{j},
\ket{-j}\}\quad.\label{hs}\ee This choice is motivated by the
continuum space canonical analysis and with the flat space
Poincare invariance. Clearly $V_j^\prime$ is a subspace of $V_j$,
and the $Spin(2)$ subgroup is generated by the $J_z$ component of
the angular momentum.

In the $j=1/2$ case we have $V(\L_s^\pm)=V_{1/2} =V^\prime_{1/2}$,
so that the only way to distinguish between the massive and a
massless fermion is by a choice of the terms in the quadratic
action (\ref{fqa}).

In the $j=1$ and $S_1 = (1/2,1/2)$ case, the index $i$ in the
massless-like $A_i$ operator will take two values, so that the
cubic interaction term would be again given by the formula
(\ref{usna}), but now all the indices will belong to the $SO(2)$
$V^\prime_\ha$ and $V^\prime_1$ representations.

Note that in the $j=1$ case one can also take $S_1^+ = (1,0)$ and
$S_1^- = (0,1)$ Lorentz irreps. These are the irreducible parts of
the antisymmetric tensor $f_{\m\n}$, and we can denote the
corresponding spin net fields as $f_i^\pm (\times)$. The spin-foam
analog of the action $\int d^4 x \bar\psi \g^{\m\n}\psi f_{\m\n}$
will be given by \be \sum_N \psi^\pm_s (\times)\psi^\pm_{s^\prime}
(\times^\prime) f^\pm_i(\times^{\prime\prime}) \cv_{\pm\pm}^{s
\,s^\prime\,i} (\times,\times^\prime,\times^{\prime\prime};
\times_1 ,...,\times_5) \f_{\times_1} \cdots \f_{\times_5} +
(h.c.) \,,\label{gic}\ee and the corresponding spin net amplitude
is given by \be A_{\pm\pm}^{s \,s^\prime\,i} = \int_{X^6} d^5 x dy
K_\ha^{st}(x_1 ,y) K_{1}^{ij}(x_4 ,y)K_\ha^{s^\prime t^\prime}
(x_2 ,y) C_{\,t \, t^\prime \, j }^{\ha\, \ha\, 1}\prod_{k<l}
K_{N_{kl}}(x_k ,x_l)\,,\label{pbpa}\ee where $K_1 = K_{(1,0),1} =
K_{(0,1),1}$.

Note that the construction (\ref{hs}) gives for $j=2$ a
massless-like helicity two field $h_{\pm 2}(\times)$. Since this
is a local excitation on a given spin net, it can be considered as
a spin foam version of the graviton. Again there are two possible
choices for the corresponding Lorentz irrep. One can take $S_2 =
(1,1)$ or $S_2^+ = (2,0)$ and $S_2^- = (0,2)$. The first choice
corresponds to a symmetric traceless tensor $h_{\m\n}$, while the
second and the third choice are the irreducible pieces of the
tensor $R_{\m\n,\r\s}$, which has the symmetry properties of the
Reimann curvature tensor. Clearly, we will then take the $(1,1)$
irrep for the graviton, and one would then use the $\tilde
K_{(1,1),2}$ propagator to construct the interaction vertices.

The realistic interactions also require an introduction of the
internal symmetry group $\tilde G$. In the framework of the spin
foam models, the simplest way to do this is to replace the
category of representations $Cat(G)$ with the category $Cat(G
\times \tilde G )$. The matter irreps would be then given by $(S
,\l_S)$, where $\l_S$ is the corresponding irrep of $\tilde G$.
The gravitational sector irreps would be $(N,1)$, where $1$ is the
trivial irrep of $\tilde G$. The matter action would again have
the form (\ref{tia}) in the topological gravity case, or
(\ref{ntia}) in the non-topological gravity case, but now the
products of the matter fields have to be contracted by the
intertwiners  $ C^{\l_1 \cdots \l_k}_{\,......}$ for the internal
group irreps.

\section{Conclusions}

The constructions we have presented are inspired by the formalism
of particle quantum field theories, and it is a straightforward
extension of the existing gravitational spin-foam models. Given
that the purely gravitational models can be related to a
discretized path-integral quantization of the BF theory
\cite{fka}, it is reasonable to expect that our model could also
be related to an analogous discretized path-integral quantization
of the BF theory with matter. One would then need to rewrite the
Einstein-Hilbert action with matter in an appropriate form, for
example \bea S_4 &=& \int \langle B \wedge R \rangle + \int
\langle
B \wedge e \wedge \bar\psi \g (\partial + \o \g ) \psi \rangle \nonumber\\
&+& \int \langle B \wedge B \rangle \left( \frac12 f^{ab}f_{ab} +
g \bar\psi \g^a \psi A_a + \cdots \right) \quad, \label{bfm}\eea
where $\o$ is the spin connection, $R= d\o + \langle \o \wedge \o
\rangle$ is the corresponding curvature two-form, $A_a = e_a^\m
A_\m$, $f_{ab} = e_a^\m e_b^\n f_{\m\n}$ and the dots stand for
other matter fields. $\langle,\rangle$ is an appropriate
contraction of the $SO(3,1)$ indices $a,b,...$, $\g$ are the
appropriate gamma matrices, $\m,\n,...$ are the curved space
indices and $e^\m_a$ are the inverse fierbeins.

Note that we have written the matter couplings in terms of the
tangent space components $A_a , ...$ . These fields carry only the
$SO(3,1)$ representation indices, and they would be the prime
candidates for the continuous space limit of our spin-network
fields. This would also mean that one needs to introduce the
fierbein form $e^a = e_\m^a dx^\m$ in the BF formalism for matter,
since it is not possible to eliminate it from the action
(\ref{bfm}) by using the constraint $B = e \wedge e$. However, in
$d=3$ case there is no such problem, since $B = e$.

As far as the graviton-like field $h(\times)$ is concerned, its
continuous space analog would be $h_{ab}(x)= e_a^\m e_b^\n
h_{\m\n}(x)$. In order to interpret this field as the graviton,
one would need to implement the constraint $h_{\m\n} = g_{\m\n} -
g^0_{\m\n}$, where $g_{\m\n}=e_\m^a e_{\n a}$ is the spacetime
metric, and $g^0_{\m\n}$ is a given background spacetime metric.
This would then imply that the amplitude for a transition from a
spatial spin net with $n$ gravitons to a spatial spin net with $m$
gravitons should not involve a sum over the background irreps $N$,
since the background is fixed. That amplitude would be then given
by the extension of the formula (\ref{fba}). Hence it is important
to study further the model, in order to better understand the
continuous spacetime limit, since there is a possibility that one
would recover the perturbative gravity scattering amplitudes in a
fixed background spacetime geometry. This also applies to the
matter fields, so that this type of study will determine the
correctness of our model.

The fact that we have taken the matter spin net fields in the
non-topological gravity model to have indices from the
representations of the $SO(3)$ and $SO(2)$ subgroups, means that
the transition amplitudes will be $SO(3)$ or $SO(2)$ group
covariant objects. One then wanders what happens with the Lorentz
covariance. There is no such problem for the spin-half fermions,
or for the $(j,0)$ and the $(0,j)$ matter irreps. However, if we
want to have the most general interactions, we need to include
$(j,k)$ irreps with $jk \ne 0$. Hence we encounter a spin-foam
analogue of the particle field theory spinor-vector interaction
problem. The only possible answer is that there should be a
spin-foam version of the gauge invariance which reduces the
Lorentz covariant expressions to $SO(3)$ or $SO(2)$ covariant
gauge-fixed expressions. In that case the expressions (\ref{ntia})
and (\ref{uone}) can be understood as gauge-fixed actions. One can
think of these actions as a spin-foam generalization of the
light-cone gauge field theory actions \cite{sig}. Actually, this
analogy with the light-cone gauge field theory is even closer,
since the finite-dimensional $SO(n)$ irreps are unitary, and hence
the corresponding spin-foam model involves only the unitary
representations, in contrast to the topological gravity spin foam
model, where the matter irreps are pseudo-unitary.

Obtaining the Lorentz covariant expressions in the non-topological
gravity case would then amount to coupling simple unitary irreps
$N$ to pseudo-unitary irreps $S$, and then forming the actions of
the type (\ref{tia}). These actions should be invariant under a
spin-foam version of the gauge transformations, so that one would
obtain the action (\ref{ntia}) in a particular gauge. A better
understanding of these issues deserves a further study.

Note that in the case of spin foam models where $A_2(\L)$ is not
$\dim\L$, it is not known what is the group field theory
formulation. However, the field theory formulation is not
necessary for the determination of the amplitudes. We used it in
order to get a physical insight for our construction. Therefore,
when $A_2 \ne dim\L$, we will postulate the amplitude formulas
(\ref{gsfa}) and (\ref{fba}), with $A_e$ and $A_v$ given by the
same expressions as in the $A_f = dim\L$ case.

An important issue which has to be explored is the question of the
perturbative finiteness of the model. As in the purely
gravitational case \cite{pez,cpr}, this will boil down to the
study of the asymptotic behaviour of the spin-net propagators
$K_N$ and $K_{S,s}$.  Note that in the Euclidian case $Tr\,
K_{S,s}$ is the same as $K_N$, where $N=(s,s)$, so that one
expects that the Euclidian spin-foam matter model should be
finite. In the Lorentzian case there is no such a simple relation
between the propagators, so that one needs to make a more detailed
analysis. Alternatively, one can formulate a quantum group version
of the construction we gave, which could be made automatically
finite if the gravitational irreps are from a finite set, and if
there are finitely many matter irreps.

The issue of the semiclassical/continuous space limit of the spin
foam models needs a more detailed study. There are no rigorous
statements about the semiclassical theory. In the purely
gravitational case one can show that in the limit of large angular
momenta one obtains amplitudes which are sums of terms
proportional to $\exp(\pm i S_R )$, where $S_R$ is the Regge form
of the Einstein-Hilbert action \cite{bw}. Actually, because of the
form of the propagator $K_N$, see (\ref{escp}) and (\ref{lscp}),
this seems to be a good approximation even for arbitrary values of
the angular momenta. Hence there are good indications that one is
on the right track. In the case of matter, one would then need to
obtain expressions containing $\exp(\pm i S_R \pm i S_{sm})$,
where $S_{sm}$ is the simplical version of the matter actions
\cite{ren}. This type of analysis could also settle the question
of which fields are really massive and which ones are massless.

\bigskip
\noindent \textbf{Acknowledgements}

I would like to thank L. Crane and R. Picken for the discussions.

\newpage

\begin{figure}[h]
\centerline{\psfig{figure=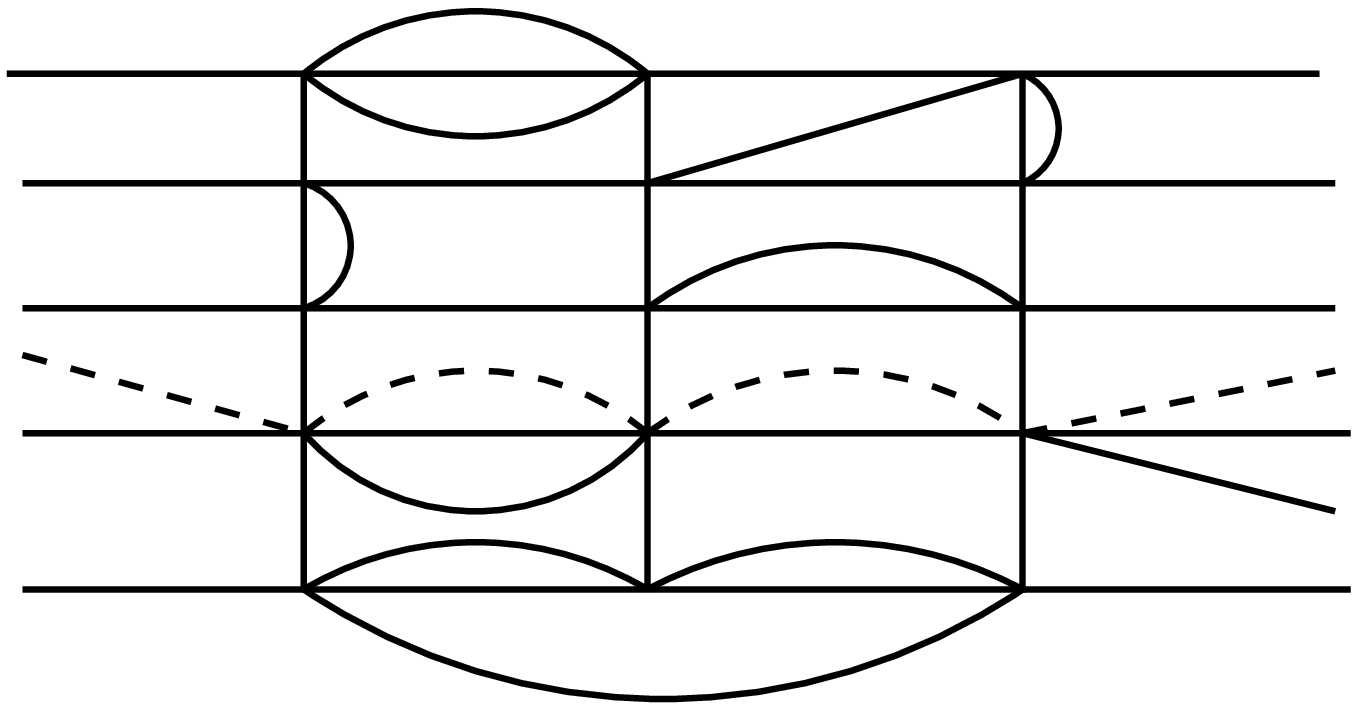}} \caption{A spin foam Feynman
diagram describing propagation of a matter field, which is
represented by a dotted line.} \label{one}
\end{figure}

\begin{figure}[h]
\centerline{\psfig{figure=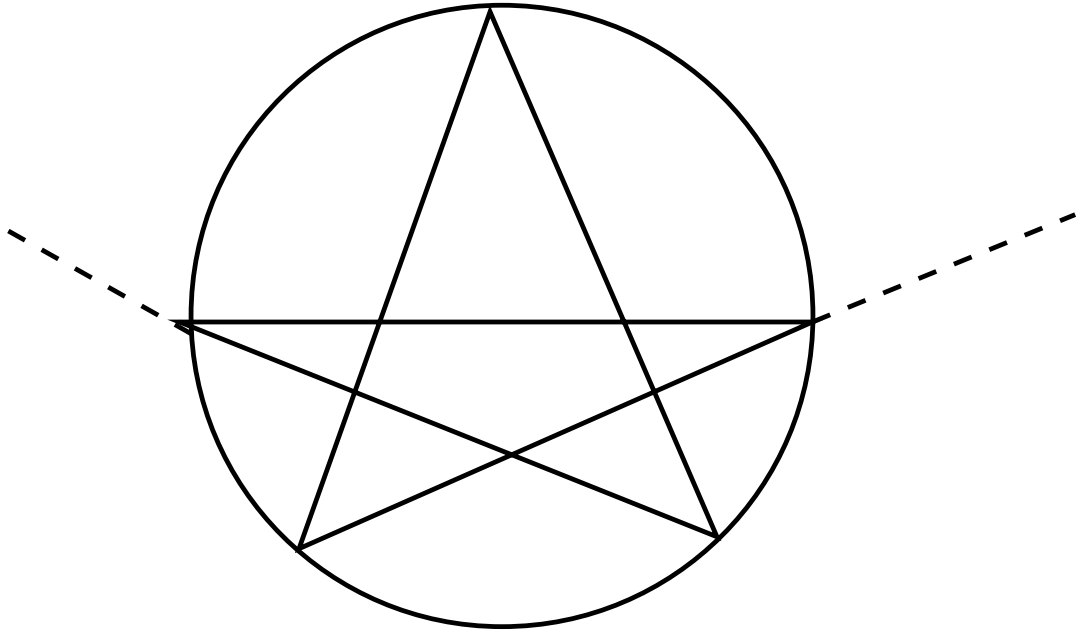}} \caption{The simplest spin net
graph which gives a 7-valent vertex for a spin foam Feynman
diagram.} \label{two}
\end{figure}

\begin{figure}[h]
\centerline{\psfig{figure=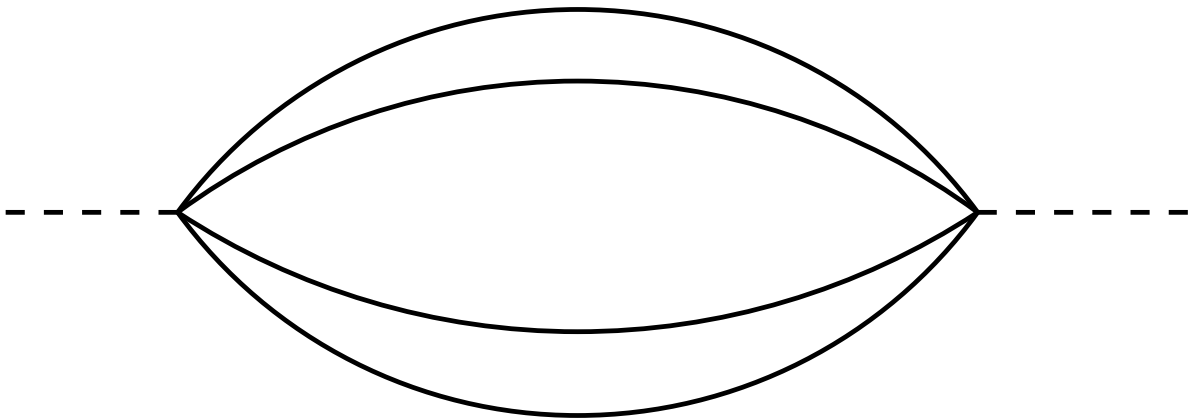}} \caption{A nontrivial spin
net graph which gives a matter propagator for a spin foam Feynman
diagram.} \label{three}
\end{figure}

\begin{figure}[h]
\centerline{\psfig{figure=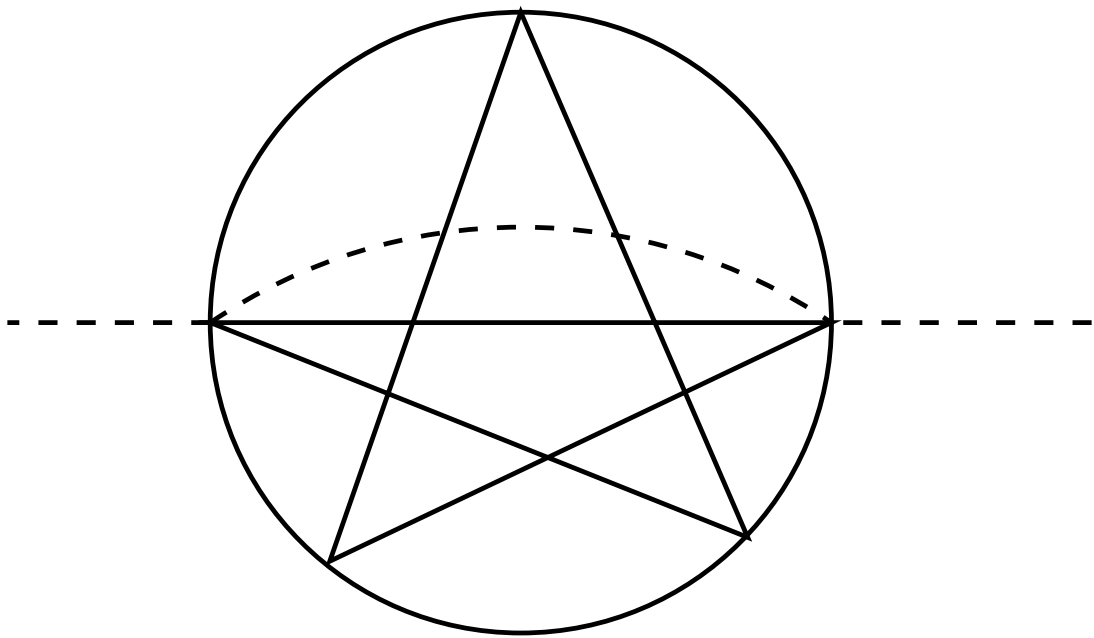}} \caption{An example of a more
complex spin net graph determining the 7-valent spin foam vertex.}
\label{four}
\end{figure}

\begin{figure}[h]
\centerline{\psfig{figure=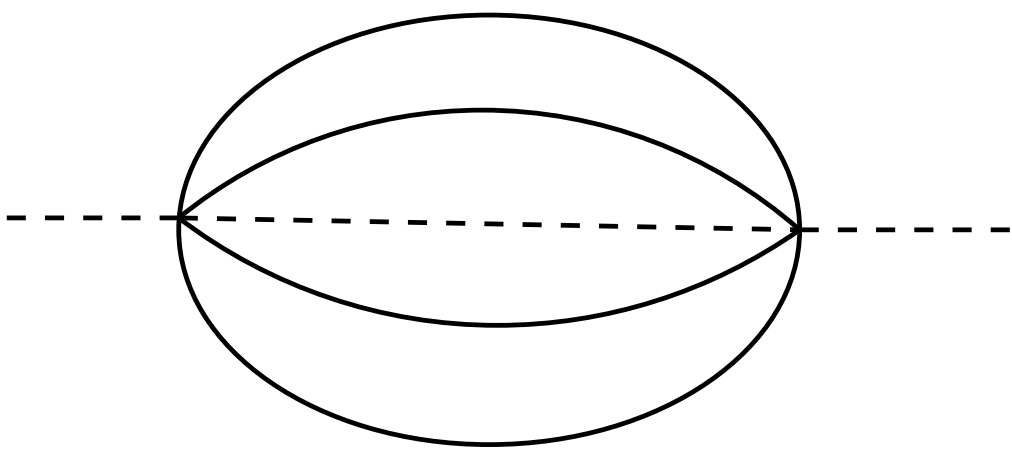}} \caption{A more complex spin
net graph determining the matter spin foam propagator.}
\label{five}
\end{figure}

\begin{figure}[h]
\centerline{\psfig{figure=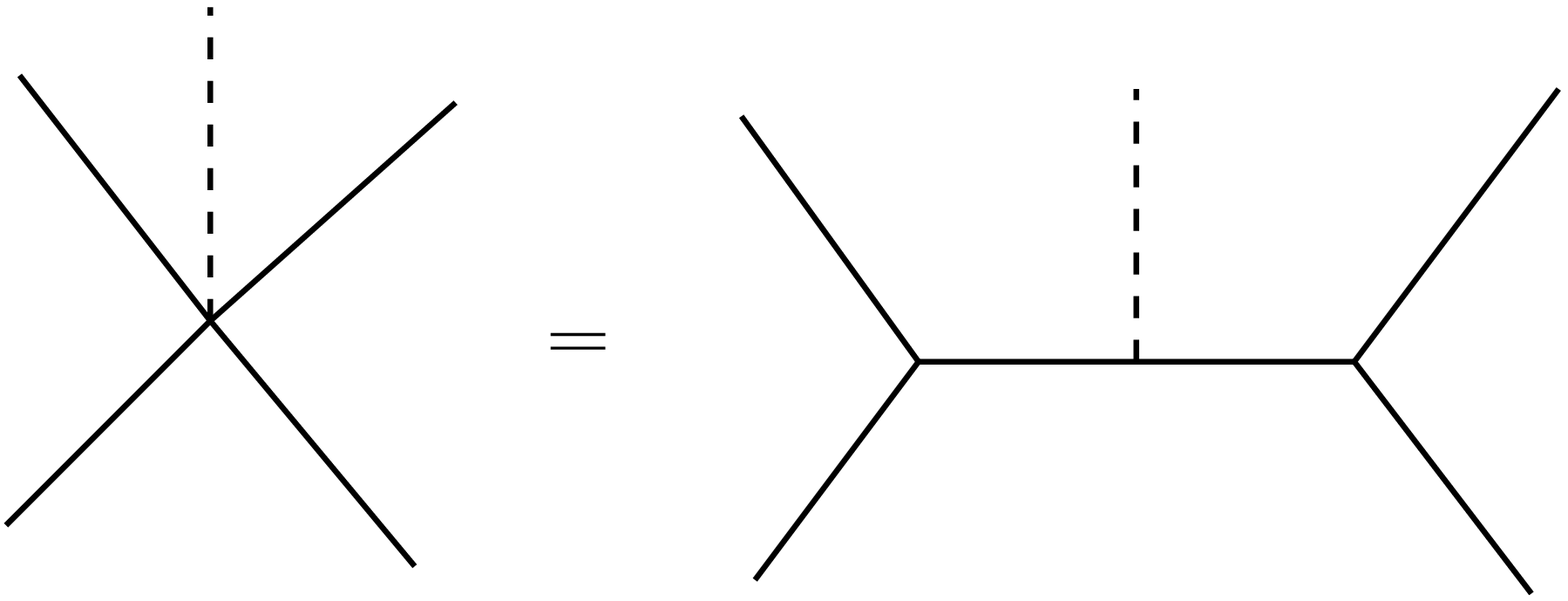}} \caption{Diagrammatic
representation of the equation (\ref{ifive}).} \label{six}
\end{figure}

\begin{figure}[h]
\centerline{\psfig{figure=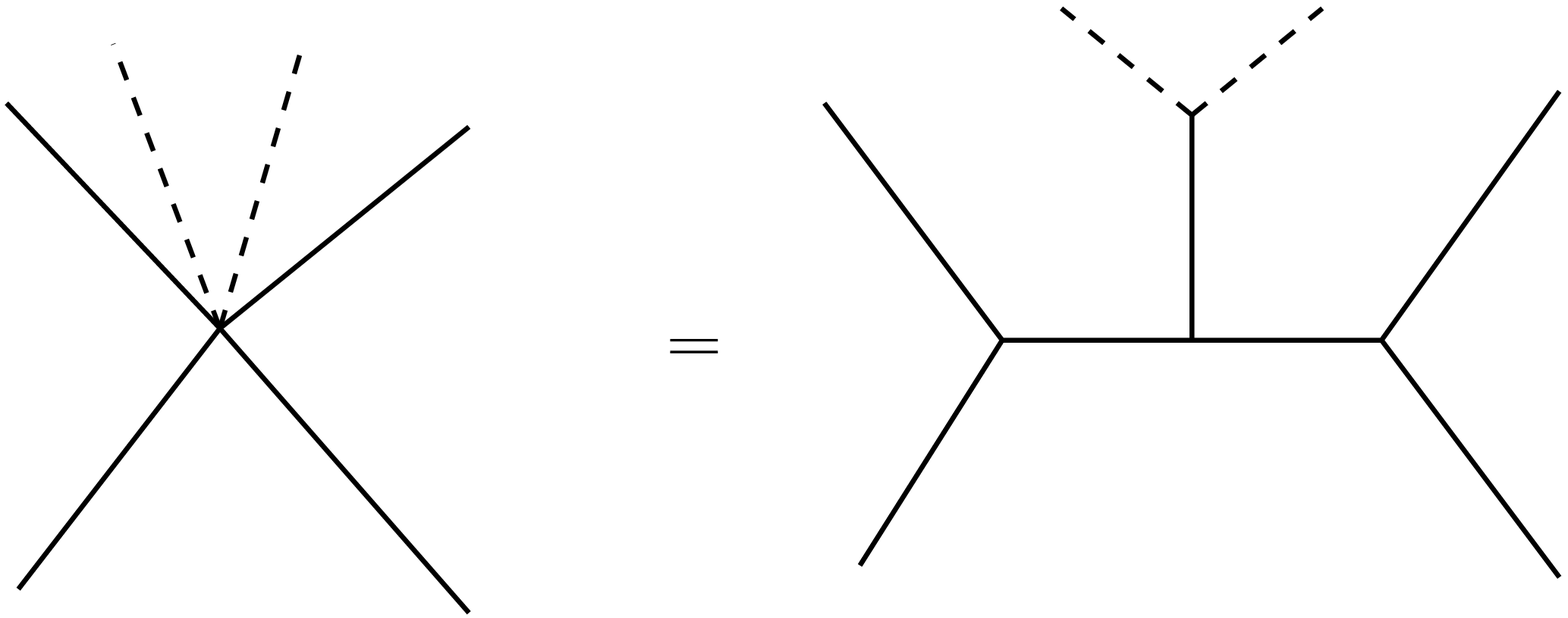}} \caption{Diagrammatic
representation of the equation (\ref{tfi}).} \label{seven}
\end{figure}

\begin{figure}[h]
\centerline{\psfig{figure=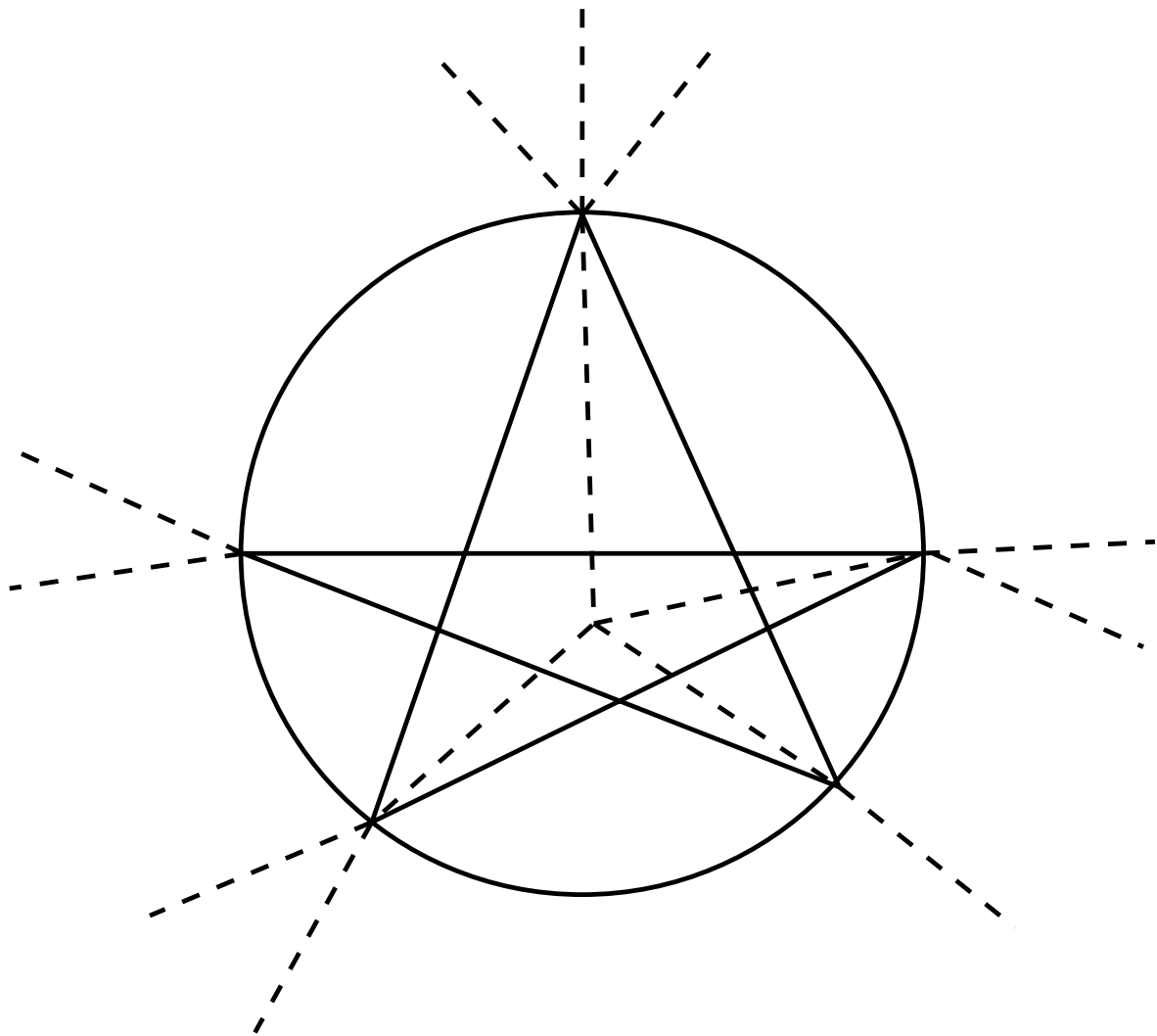}} \caption{A spin net graph
determining a 15-valent spin foam vertex.} \label{nine}
\end{figure}

\begin{figure}[h]
\centerline{\psfig{figure=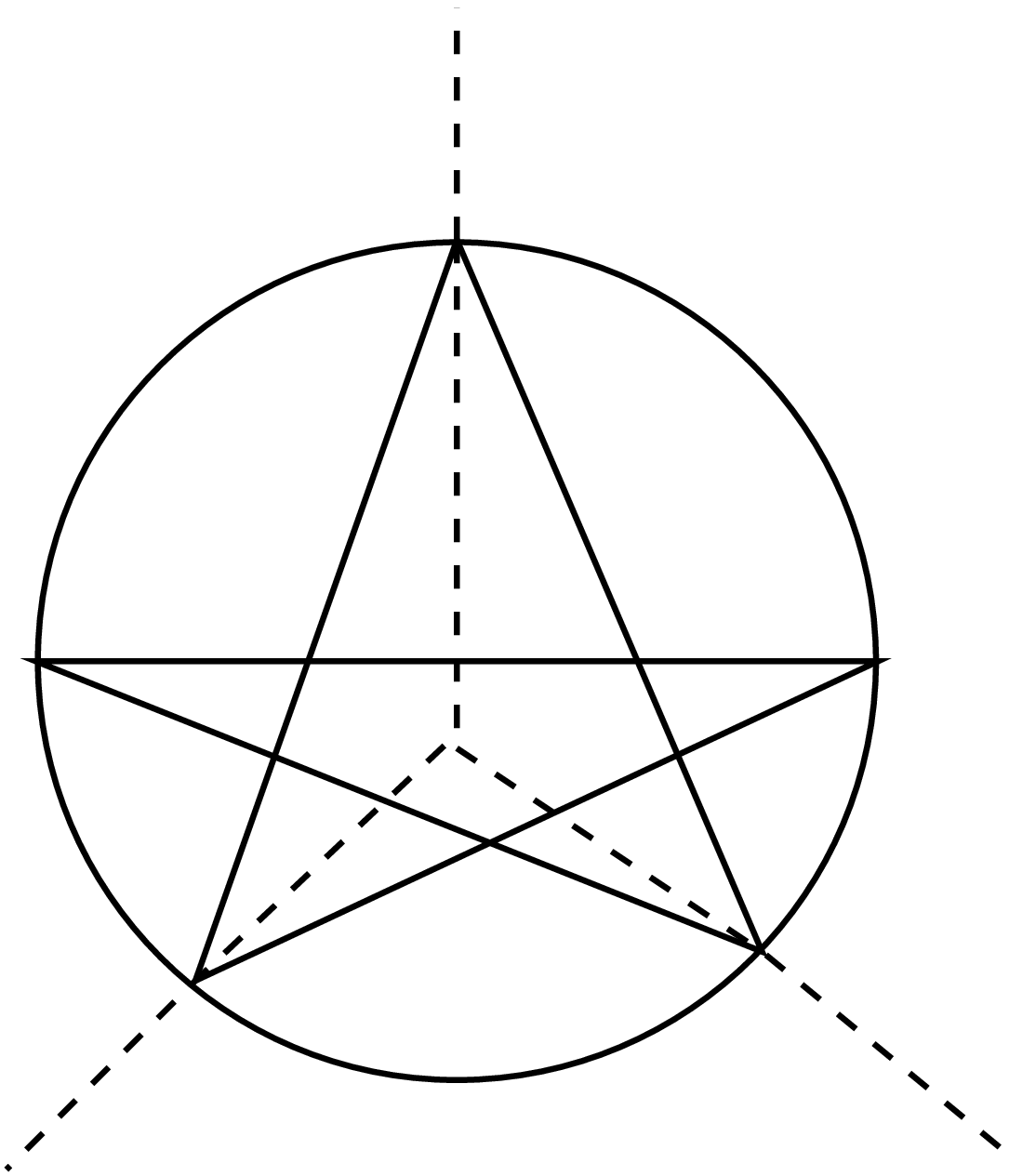}} \caption{The spin net graph
for the spin foam vector-spinor interaction vertex.} \label{ten}
\end{figure}


\begin{thebibliography}{99}

\bibitem{bce} J.W. Barrett and L. Crane, J. Math. Phys. 39 (1998) 3296
\bibitem{bcl} J.W. Barrett and L. Crane, Class. Quant. Grav. 17 (2000) 3101
\bibitem{pre} A. Perez and C. Rovelli, Nucl. Phys. B 599 (2001) 255
\bibitem{prl} A. Perez and C. Rovelli, Phys. Rev. D 63 (2001) 041501
\bibitem{mik} A. Mikovi\'c, Class. Quant. Grav. 18 (2001) 2827

\bibitem{tv} V.G. Turaeev and O.Y. Viro, Topology 31 (1992) 865
\bibitem{cky} L. Crane, D.H. Kauffman and D.N. Yetter, J. Knot Th. Ramif.
6 (1997) 177, hep-th/9409167

\bibitem{boul} D.V. Boulatov, Mod. Phys. Lett. A7 (1992) 1629
\bibitem{oo} H. Ooguri, Mod. Phys. Lett. A7 (1992) 2799
\bibitem{dfkr} R. De Pietri, L. Freidel, K. Krasnov and C. Rovelli,
Nucl. Phys. B 574 (2000) 785
\bibitem{rr} M. Reisenberger and C. Rovelli, Class. Quant. Grav.
18 (2001) 121
\bibitem{lpr} R. Levine, A. Perez, C. Rovelli, gr-qc/0102051

\bibitem{pez} A. Perez, Nucl. Phys. B 599 (2001) 427
\bibitem{cpr} L. Crane, A. Perez and C. Rovelli, Phys. Rev. Lett.
87 (2001) 181301

\bibitem{cr} L. Crane, gr-qc/0004043
\bibitem{c2} L. Crane, gr-qc/0110060

\bibitem{w} S. Weinberg, The Quantum Theory of Fields, vol. 1,
Cambridge University Press (1995) Cambridge

\bibitem{rs} C. Rovelli and L. Smollin, Phys. Rev. D 52 (1995) 5743

\bibitem{op} R. Oeckl and H. Pfeiffer, Nucl. Phys. B 598 (2001) 400
\bibitem{pf} H. Pfeiffer, hep-th/0106029

\bibitem{mtr}H.A. Morales-Tecotl and C. Rovelli, Phys. Rev. Lett. 72 (1994) 3642
\bibitem{bk}J. Baez and K. Krasnov, J. Math. Phys. 39 (1998) 1251

\bibitem{ba}J. Baez, Lect. Notes Phys. 543 (2000) 25
\bibitem{or} D. Oriti, Rept. Prog. Phys. 64 (2001) 1489


\bibitem{bar} J.W. Barrett, Adv. Theor. Math. Phys. 2 (1998) 593
\bibitem{fkb}L. Freidel and K. Krasnov, J. Math. Phys. 41 (2000) 1681

\bibitem{ch} R. Camporesi and A. Higuchi, J. Geom. Phys. 15 (1996) 1, gr-qc/9505009

\bibitem{ruhl} W. Ruhl, The Lorentz Group and Harmonic Analysis,
WA Benjamin Inc (1970) New York


\bibitem{fka} L. Freidel and K. Krasnov, Adv. Theor. Math. Phys. 2 (1999) 1183

\bibitem{sig} W. Siegel, Introduction to String Field theory,
World Scientific (1988) Singapoore, hep-th/0107094

\bibitem{bw} J.W. Barrett and R.M. Williams, Adv. Theor. Math.
Phys. 3 (1999) 1

\bibitem{ren} H-C. Ren, Nucl. Phys. B 301 (1988) 661

\end{thebibliography}
\end{document}